\documentclass[aps,12pt,a4paper,amsfonts,amssymb,titlepage,showpacs,preprint,showkeys,nofootinbib,nobibnotes,byrevtex,superscriptaddress]{revtex4}%
\usepackage{amsfonts}
\usepackage{amsmath}
\usepackage{amssymb}
\usepackage{graphicx}%
\providecommand{\U}[1]{\protect\rule{.1in}{.1in}}
\begin{document}
\title{Harada-Tsutsui Gauge Recovery Procedure: From Abelian Gauge Anomalies to the Stueckelberg Mechanism}
\author{Gabriel Di Lemos Santiago Lima}
\email{gsl11@psu.edu, gabriellemos3@hotmail.com}
\affiliation{Institute for Gravitation and the Cosmos, and Physics Department, 
The Pennsylvania State University, University Park, PA 16802, USA}

\keywords{gauge field theories; gauge anomalies; nonperturbative techniques }
\pacs{11.15.-q; 11.15.Tk; 11.30.-j}

\begin{abstract}
Revisiting a path-integral procedure of recovering gauge invariance 
from anomalous effective actions developed by Harada and Tsutsui, it is shown that there are two ways to achieve
gauge symmetry: one already presented by the authors,
which is shown to preserve the anomaly in the sense of standard 
current conservation law, and another one which is anomaly-free, 
preserving current conservation. It is also shown that the application 
of the Harada-Tsutsui technique to other models which are not anomalous
but do not exhibit gauge invariance allows the identification of the 
gauge invariant formulation of the Proca model, also done by the 
referred authors, with the Stueckelberg model, leading to the 
interpretation of the gauge invariant map as a generalization of the 
Stueckelberg mechanism. 
 
\end{abstract}
\maketitle
\section{Introduction}
A gauge anomalous theory is one that presents a breakdown of gauge 
invariance at the quantum level \cite{J}. When it happens, it is shown that 
the expectation value of the current divergence is not identically null, but, 
instead, there remains a term which is a function of the gauge field, which 
is called the anomaly. In this sense, it is used to be widely believed that there is a breaking of the 
current conservation law due to the presence of the gauge anomaly. This is one 
of the reasons why these models are not well liked, besides problems with 
renormalizability due to non-gauge invariance.

There is a range of contexts 
in which the discussion around anomalies is brought up, such as superstrings \cite{Polch}, quantum gravity \cite{WG,RW} and 
condensed matter phenomena description such as the fractional quantum 
Hall effect \cite{GT}, for example. One of the fundamental prerequisites to renormalization and unitarity is the existence of the 
Stanislaw-Taylor identities, which seems to be spoiled by the presence of a 
gauge anomaly. For instance, one of the most important discussion was played by theories of Weyl 
fermions coupled to gauge fields, where the appearance of gauge anomalies is viewed as 
unavoidable, due to its quantum competition with chiral anomalies \cite{comp}. 
However, recently it was shown that when one goes to the full quantum level, 
where the gauge field is also quantized, then the expectation value of the 
anomaly must vanish \cite{GRT}. For these reasons, it may be worth to re-discuss 
this subject in more detail.  

In the eighties, an amount of discussion about anomalous models in
quantum field theory was presented. The central role of discussion 
was played by consistence of such theories. Although some theorists
considered such models as inconsistents, some authors produced 
works to support the idea that they are not so.

In this sense, we must cite the work of Jackiw and Rajaraman \cite{JR},
in which it was shown that a gauge anomalous two-dimensional theory
could be well defined and be able to provide a mass generation 
mechanism from chiral anomalies. This work was soon followed by
the one of Fadeev and Shatashvili \cite{FS}, who noticed that quantum
gauge invariance could be restored by the introduction of new degrees
of freedom that transform second class constraints into first class ones.
In adding these extra fields, the effective anomalous action is mapped
into a gauge invariant one. Then, the works of Babelon, Schaposnick
and Viallet \cite{BSV} and Harada and Tsutsui \cite{HT} showed 
independently that these degrees of freedom could emerge quite 
naturally by the application of Faddeev-Popov's method through the 
non-factorization of the integration over the gauge group. Soon after,
Harada and Tsutsui recognized that the same procedure could be applied
to the Proca model \cite{HT2}, leading to  possible generalization of their
technique. 

We can recognize the main strategy to give consistence to these models 
with the introduction of the new degrees of freedom, which recovers gauge
invariance. In this sense, it seems useful to analyze such procedure and 
explore its potential. Although restoring gauge symmetry at the final 
effective action, one may ask whether such technique is able to provide
current conservation or it just preserve the quantum anomaly.

This work is intended to elucidate this question for the particular case of
abelian gauge anomalies. In this sense, in section II the origin of abelian
gauge anomaly is briefly reviewed in path integral approach. In section
III, the gauge invariant formalism developed by Harada and Tsutsui (HT) is
rederived by redefining the vacuum functional multiplying it by the gauge
volume, instead of proceeding with Faddeev-Popov's method, and it is 
shown that the anomaly is preserved in the original form proposed by 
the authors. Section IV is intended to show that their procedure gives 
rise to another abelian gauge invariant formulation which may provide
an anomaly free model. In section V, the HT procedure
applied to the Proca model is rederived. Finally, in section VI, a 
correspondence between the Proca's gauge invariant mapping and the 
Stueckelberg model is pointed out, leading to the interpretation of the
HT procedure as a generalization of the Stueckelberg 
mechanism \cite{Stueck}. The conclusion is, then, presented in section VII.
 
\section{The Origin of Abelian Gauge Anomaly in Path Integral Approach}

Consider an abelian gauge theory described by the action
\begin{eqnarray}
I[\psi,\overline{\psi},A_{\mu}]  &  =I_{M}[\psi,\overline{\psi},A_{\mu}]+I_{G}[A_{\mu}] 
\end{eqnarray}
where $I_{M}[\psi,\overline{\psi},A_{\mu}]$ is the matter action minimally 
coupled to the abelian gauge field $A_{\mu}$, and $I_{G}[A_{\mu}]$ is the free 
bosonic action. If the action is said to be invariant under local gauge 
transformations
\begin{align}
\psi\rightarrow\psi^{\theta}&=exp(i\theta(x))\psi\\
\bar{\psi}\rightarrow\bar{\psi}^{\theta}&=exp(-i\theta(x))\bar{\psi}\\
A_{\mu}\rightarrow A_{\mu}^{\theta}&=A_{\mu}+\frac{1}{e}\partial_{\mu}\theta(x),
\end{align}
one can say that, classically, the theory exhibits a conserved current given by
\begin{equation}
  J^{\mu}=-\frac{1}{e}\frac{\delta{I_{M}}}{\delta{A_{\mu}}}.
\end{equation}
Now, if we proceed the quantization of the fermionic fields, then, after 
integrating them out, we will arrive at an effective action give by
\begin{equation}
\exp(iW[A])=\int d\psi d\bar{\psi}\exp(iI[\psi,\bar{\psi},A]). \label{eff-action}
\end{equation}

To find the quantum version of the current conservation law, first we make a 
change of variables in the fermion fields
\begin{eqnarray}
\exp(iW[A])&=&\int d\psi d\bar{\psi}\exp(iI[\psi,\bar{\psi},A])\nonumber\\
&=&\int d\psi^{\theta} 
d\bar{\psi}^{\theta}\exp(iI[\psi^{\theta},\bar{\psi}^{\theta},A]),
\end{eqnarray}
and then, just as in classical case, we make use of the invariance of the action 
by noticing that $I[\psi^{\theta},\bar{\psi}^{\theta},A]=I[\psi,\bar{\psi},A^{-\theta}]$
\begin{equation}
\exp(iW[A])=\int d{\psi}^{\theta} 
d\bar{\psi}^{\theta}\exp(iI[\psi,\bar{\psi},A^{-\theta}])
\end{equation}
Now, a subtle difference between the classical and the quantum gauge theory 
arises: if the quantum measure is \textit{locally} gauge invariant, \textit{i. e.}, if
\begin{equation}
d\psi d\bar{\psi}=d\psi^{\theta} d\bar{\psi}^{\theta},  \label{fmeasure}
\end{equation}
then, by considering $\theta(x)$ as an infinitesimal parameter, we will have
\begin{eqnarray}
\exp(iW[A])&=&\int d{\psi}^{\theta} 
d\bar{\psi}^{\theta}\exp\left(iI\left[\psi,\bar{\psi},A^{-\theta}\right]\right)\nonumber\\
&=&\int 
d{\psi}d\bar{\psi}\exp\left(iI\left[\psi,\bar{\psi},A_{\mu}-\frac{1}{e}\partial_{\mu}\theta(x)\right]\right)\nonumber\\
&=&\exp(iW[A])-\int dx i\theta(x)\int d\psi d\bar{\psi}\partial_{\mu}\left(-\frac{1}{e}\frac{\delta{I}}{\delta{A_{\mu}}}\right)\exp(iI[\psi,\bar{\psi},A_{\mu}])
\end{eqnarray}
\begin{equation}
\Rightarrow\int d{\psi}d\bar{\psi}\partial_{\mu}\left(-\frac{1}{e}\frac{\delta 
I}{\delta{A_{\mu}}}\right)\exp(iI[\psi,\bar{\psi},A_{\mu}])=0.
\end{equation}
But gauge invariance of the free bosonic action implies that $\partial_{\mu}\left(\frac{\delta I_{G}}{\delta{A_{\mu}}}\right)\equiv 
0$, therefore,
\begin{equation}
\int d{\psi}d\bar{\psi}\partial_{\mu}\left(-\frac{1}{e}\frac{\delta 
I_{M}}{\delta{A_{\mu}}}\right)\exp(iI[\psi,\bar{\psi},A_{\mu}])=0. \label{qconserv-curr}
\end{equation} 

Equation (\ref{qconserv-curr}) is the quantum version of the current conservation law. 
However, it was necessary to impose invariance of the fermionic measure (\ref{fmeasure}) 
to get the above result. If, instead of (\ref{fmeasure}), we had
\begin{equation}
d\psi^{\theta}d\bar{\psi}^{\theta}=\exp\left(i\alpha_{1}[A,\theta]\right)d\psi d\bar{\psi} 
\label{wess-zumino}
\end{equation}
then, instead of (\ref{qconserv-curr}), we would arrive at
\begin{eqnarray}
\exp(iW[A])&=&\int d\psi^{\theta} 
d\bar{\psi}^{\theta}\exp\left(iI\left[\psi,\bar{\psi},A^{-\theta}\right]\right)\nonumber\\
&=&\int d\psi 
d\bar{\psi}\exp\left(iI\left[\psi,\bar{\psi},A^{-\theta}\right]+i\alpha_{1}[A,\theta]\right)\nonumber\\
&=&\int d\psi 
d\bar{\psi}\exp\left\{iI\left[\psi,\bar{\psi},A\right]
+i\int dx\partial_{\mu}\theta(x)\left(-\frac{1}{e}\frac{\delta{I}}{\delta 
A_{\mu}}\right)\right. \nonumber\\
&+&\left. \left. \alpha_{1}[A,0]+i\int dx\frac{\delta\alpha_{1}}{\delta\theta}\right|_{\theta=0}\theta(x)\right\}\nonumber,
\end{eqnarray}
but $\partial_{\mu}\left(-\frac{1}{e}\frac{\delta I}{\delta A_{\mu}}\right)
=\partial_{\mu}\left(-\frac{1}{e}\frac{\delta I_{M}}{\delta A_{\mu}}\right)$ and
$\alpha_{1}[A,0]=0$, therefore
\begin{eqnarray}
\exp(iW[A])&=&\int d\psi d\bar{\psi}\exp\left\{iI\left[\psi,\bar{\psi},A\right]-i\int dx\theta(x)\left[\partial_{\mu}\left(-\frac{1}{e}\frac{\delta I_{M}}{\delta 
A_{\mu}}\right)-\left. \frac{\delta\alpha_{1}}{\delta\theta}\right|_{\theta=0}\right]\right\}\nonumber\\
&=&\int d\psi d\bar{\psi}\exp\left(iI\left[\psi,\bar{\psi},A\right]\right)\left\{1-i\int dx\theta(x)\left[\partial_{\mu}\left(-\frac{1}{e}\frac{\delta I_{M}}{\delta 
A_{\mu}}\right)-\left. \frac{\delta\alpha_{1}}{\delta\theta}\right|_{\theta=0}\right]\right\}\nonumber\\
&=&\exp(iW[A])-i\int dx\theta(x)\int d\psi d\bar{\psi}\exp\left(iI\left[\psi,\bar{\psi},A_{\mu}\right]\right)\left[\partial_{\mu}\left(-\frac{1}{e}\frac{\delta I_{M}}{\delta 
A_{\mu}}\right)-\left. \frac{\delta\alpha_{1}}{\delta\theta}\right|_{\theta=0}\right]\nonumber
\end{eqnarray}
\begin{equation}
\Rightarrow\int d\psi d\bar{\psi}\partial_{\mu}\left(-\frac{1}{e}\frac{\delta I_{M}}{\delta 
A_{\mu}}\right)\exp\left(iI\left[\psi,\bar{\psi},A_{\mu}\right]\right)=\mathcal{A}\exp(iW[A]), \label{anoeq}
\end{equation}
and we see that, instead of (\ref{qconserv-curr}), we would have a non-vanishing 
right-hand side in (\ref{anoeq}), were
\begin{equation}
\mathcal{A}\equiv\left. \frac{\delta\alpha_{1}}{\delta\theta}\right|_{\theta=0} \label{ano}
\end{equation}
is called the anomaly and the theory is said to be anomalous.

It is convenient, to our purposes, to rewrite the anomaly (\ref{ano}) by 
noticing that
\begin{eqnarray}
\left. \frac{\delta\alpha_{1}}{\delta\theta}\right|_{\theta=0}&=&\left. \frac{\delta 
W\left[A^{\theta}\right]}{\delta\theta}\right|_{\theta=0}\nonumber\\
&=&\int d^{n}x\left(\frac{1}{e}\frac{\delta W[A]}{\delta 
A_{\mu}(y)}\right)\partial_{\mu}[\delta (x-y)]\nonumber\\
&=&\partial_{\mu}\left(\frac{1}{e}\frac{\delta W[A]}{\delta 
A_{\mu}(y)}\right),\nonumber
\end{eqnarray}
and, therefore
\begin{equation}
\mathcal{A}\equiv\left. \frac{\delta\alpha_{1}}{\delta\theta}\right|_{\theta=0}=\partial_{\mu}\left(\frac{1}{e}\frac{\delta W[A]}{\delta 
A_{\mu}(x)}\right).
\end{equation}

\section{Gauge Invariant Formulation of Anomalous Models}

The anomaly arises from the non-invariance of the effective action. To see this, 
we notice that
\begin{eqnarray}
\exp(iW\left[A^{\theta}\right])&=&\int d\psi 
d\bar{\psi}\exp\left(iI\left[\psi,\bar{\psi},A^{\theta}\right]\right)\nonumber\\
&=&\int d\psi^{\theta} 
d\bar{\psi}^{\theta}\exp\left(iI\left[\psi^{\theta},\bar{\psi}^{\theta},A^{\theta}\right]\right)\nonumber\\
&=&\int d\psi 
d\bar{\psi}\exp\left(iI\left[\psi,bar{\psi},A^{\theta}\right]+i\alpha_{1}[A,\theta]\right)\nonumber\\
&=&\exp\left(iW[A]+i\alpha_{1}[A,\theta]\right),
\end{eqnarray}
that is,
\begin{equation}
\Rightarrow\alpha_{1}[A,\theta]=W\left[A^{\theta}\right]-W[A].  
\end{equation}
Therefore, from (\ref{anoeq}) it seems that current conservation at quantum 
level may be obtained only for theories with gauge invariant effective actions.

A gauge invariant formulation of anomalous theories was built by Harada and 
Tsutsui in \cite{HT}. We will derive the same results in a different way that 
is more convenient to our purposes, instead of inserting the usual Faddeev-Popov identity. It is considered the \textit{full} theory, described 
by the vacuum functional
\begin{eqnarray}
 Z&=&\int d\psi d\bar{\psi}dA_{\mu}\exp(iI[\psi,\bar{\psi},A])\nonumber\\
 &=&\int dA_{\mu}\exp(iW[A]).
\end{eqnarray}
The functional can be redefined by multiplying it by the gauge volume and, then, 
a change of variables in the gauge field can be performed
\begin{eqnarray}
 Z&=&\int d\theta dA_{\mu}\exp(iW[A])\nonumber\\
 &=&\int d\theta dA_{\mu}^{\theta}\exp(iW\left[A^{\theta}\right]).
\end{eqnarray}
Now we use the fact that the boson measure \textit{is} gauge invariant, that is 
$dA_{\mu}=dA_{\mu}^{\theta}$, and we arrive at a theory containing a scalar 
field $\theta$, besides the gauge field $A_{\mu}$
\begin{eqnarray}
 Z&=&\int d\theta dA_{\mu}\exp(iW'[A,\theta])\nonumber\\
 &=&\int dA_{\mu}\exp(iW_{eff}\left[A\right]), \label{eff-funct}
\end{eqnarray}
where
\begin{equation}
W'[A,\theta]\equiv W\left[A^{\theta}\right] \text{and} 
\exp\left(iW_{eff}[A]\right)\equiv \int d\theta \left(iW'[A,\theta]\right) \label{eff-theo}
\end{equation}

It is easy to see that the new effective action $W_{eff}[A]$ is gauge invariant. 
To do this, we notice that
\begin{eqnarray}
\exp\left(iW_{eff}\left[A^{\lambda}\right]\right)&=&\int 
d\theta\exp\left(iW'\left[A^{\lambda},\theta\right]\right)\nonumber\\
&=&\int d\theta\exp\left(iW'\left[A,\theta+\lambda\right]\right)\nonumber\\
&=&\int d(\theta+\lambda)\exp\left(iW'\left[A,\theta+\lambda\right]\right)\nonumber\\
&=&\exp\left(iW_{eff}[A]\right).
\end{eqnarray}

One could ask if, after this procedure, the anomaly would survive, and we can 
say that it depends on the starting action. Indeed, one may choose an initial 
action by noticing that
\begin{eqnarray}
 Z&=&\int d\theta dA_{\mu}\exp\left(iW'[A,\theta]\right)\nonumber\\
 &=&\int d\theta dA_{\mu}\exp\left(iW\left[A^{\theta}\right]\right)\nonumber\\
 &=&\int d\theta dA_{\mu}\exp\left(iW[A]+i\alpha_{1}[A,\theta]\right)\nonumber\\
 &=&\int d\theta dA_{\mu}\exp\left(iI\left[\psi,\bar{\psi},A\right]+i\alpha_{1}[A,\theta]\right). \label{st.act}
\end{eqnarray}

The action in eq. (\ref{st.act}), with the addition of the Wess-Zumino term $\alpha_{1}[A,\theta]$ 
\cite{WZ}, is known as the standard action \cite{HT}
\begin{equation}
I_{st}\left[\psi,\bar{\psi},A,\theta\right]=I\left[\psi,\bar{\psi},A\right]+\alpha_{1}[A,\theta]. \label{st.act'}
\end{equation}

As one could notice, although the final effective action $W_{eff}[A]$ is gauge 
invariant, the standard one $I_{st}\left[\psi,\bar{\psi},A,\theta\right]$ 
\textit{is not}, since $\alpha_{1}[A,\theta]$ breaks gauge invariance. To 
understand what it means, we see that, if we search for a kind of conserved 
current from this theory, we need to start from the gauge invariance of the 
effective action, which leads to
\begin{equation}
\partial_{\mu}\left(-\frac{1}{e}\frac{\delta W_{eff}[A]}{\delta A_{\mu}(x)}\right)=0.  
\end{equation}
Then we have
\begin{eqnarray}
&\partial_{\mu}&\left(-\frac{1}{e}\frac{\delta W_{eff}[A]}{\delta 
A_{\mu}(x)}\right)\nonumber\\
&=&\frac{i}{e}\partial_{\mu}\left\{\frac{\delta}{\delta A_{\mu}(x)}\exp\left(iW_{eff}[A]\right)\right\}\nonumber\\
&=&\frac{i}{e}\partial_{\mu}\left\{\frac{\delta}{\delta A_{\mu}(x)}\int d\theta d\psi d\bar{\psi}\exp\left(iI_{st}[\psi,\bar{\psi},A,\theta]\right)\right\}\nonumber\\ 
&=&\int d\theta d\psi d\bar{\psi}\partial_{\mu}\left(-\frac{i}{e}\frac{\delta I_{st}}{\delta A_{\mu}(x)}\right)\exp\left(iI_{st}[\psi,\bar{\psi},A,\theta]\right)\nonumber\\   
&=&\int d\theta d\psi d\bar{\psi}\partial_{\mu}\left(-\frac{i}{e}\frac{\delta I_{M}\left[\psi,\bar{\psi},,A\right]}{\delta A_{\mu}(x)}--\frac{i}{e}\frac{\delta \alpha_{1}[A,\theta]}{\delta A_{\mu}(x)}\right)\exp\left(iI_{st}[\psi,\bar{\psi},A,\theta]\right)=0,   
\end{eqnarray}
and since $\alpha_{1}[A,\theta]$ is not gauge invariant, one cannot say that $\partial_{\mu}\left(-\frac{1}{e}\frac{\delta\alpha_{1}[A,\theta]}{\delta 
A_{\mu}(x)}\right)=0$, which would lead to the current conservation law. Instead, we 
have
\begin{eqnarray}
\int &d\theta& d\psi 
d\bar{\psi}\partial_{\mu}J^{\mu}\exp\left(iI_{st}\left[\psi,\bar{\psi},A,\theta\right]\right)\nonumber\\
=\int &d\theta& d\psi d\bar{\psi}\partial_{\mu}\left(-\frac{1}{e}\frac{\delta\alpha_{1}[A,\theta]}{\delta 
A_{\mu}(x)})\right)\exp\left(iI_{st}\left[\psi,\bar{\psi},A,\theta\right]\right)\neq 0. \label{non-consev-curr}
\end{eqnarray}

Now, we can perform integration over the $\theta$ field in the right-hand side 
of (\ref{non-consev-curr}), using (\ref{wess-zumino}), (\ref{eff-action}) and 
the gauge invariance of $W_{eff}[A]$. It is straightforward to find
\begin{equation}
\int d\theta d\psi 
d\bar{\psi}\partial_{\mu}J^{\mu}\exp\left(iI_{st}\left[\psi,\bar{\psi},A,\theta\right]\right)=\mathcal{A}\exp\left(iW_{eff}[A]\right),  
\end{equation}
and we see that the standard formulation still preserves the anomaly, in spite of 
being invariant at the effective theory. This may be explained by the switching of gauge symmetry breakdown from the effective action to the starting one, namely, the standard action. 

\section{Recovering Current Conservation} 

The standard action is not the only one that can provide the gauge invariant 
effective theory given by (\ref{eff-theo}). Indeed, from (\ref{eff-funct}) we 
have
\begin{eqnarray}
Z&=&\int d\theta dA_{\mu}\exp(iW'[A,\theta])\nonumber\\
&=&\int d\theta dA_{\mu}\exp\left(iW[A^{\theta}]\right)\nonumber\\
&=&\int d\theta d\psi d\bar{\psi} dA_{\mu}\exp\left(iI\left[\psi,\bar{\psi},A^{\theta}\right]\right).  
\end{eqnarray}
Thus, we can see that the same procedure that leads to (\ref{eff-funct}) and (\ref{eff-action}) 
can be done by a rather more obvious choice
\begin{equation}
I_{en}[\psi,\bar{\psi},A,\theta]\equiv I[\psi,\bar{\psi},A^{\theta}].
\end{equation}
which we can call, just to distinguish from the standard action, the \textit{enhanced} action.

The advantage of this action is that it is really gauge invariant. Moreover, if we start from the gauge invariance of $W_{eff}[A]$ and proceed the same calculations which lead to (\ref{non-consev-curr}), we will arrive at
\begin{eqnarray}
&\partial_{\mu}&\left(-\frac{1}{e}\frac{\delta W_{eff}[A]}{\delta A_{\mu}(x)}\right)\exp(iW_{eff}[A])\nonumber\\ 
&=&\int d\theta d\psi d\bar{\psi}\partial_{\mu}\left(-\frac{1}{e}\frac{\delta I_{en}}{\delta A_{\mu}(x)}\right)\exp\left(I_{en}[\psi,\bar{\psi},A,\theta]\right)\nonumber\\ 
&=&\int d\theta d\psi d\bar{\psi}\partial_{\mu}\left(-\frac{1}{e}\frac{\delta I\left[\psi,\bar{\psi},A^{\theta}\right]}{\delta A_{\mu}(x)}\right)\exp\left(I_{en}[\psi,\bar{\psi},A,\theta]\right)=0 \label{en-curr}
\end{eqnarray}

In fermionic theories, generally the gauge fields are coupled linearly to the 
fermions. So, expanding the matter action to the first order, we will obtain
\begin{eqnarray}
I\left[\psi,\bar{\psi},A\right]&=&I_{M}\left[\psi,\bar{\psi},A\right]+I_{G}[A]\nonumber\\
&=&I_{F}\left[\psi,\bar{\psi}\right]+\int d^{n}x\frac{\delta I_{M}\left[\psi,\bar{\psi},A\right]}{\delta A_{\mu}(x)}+I_{G}[A], 
\end{eqnarray}
where $I_{F}\left[\psi,\bar{\psi}\right]\equiv I_{M}\left[\psi,\bar{\psi},0\right]$ 
corresponds to the free fermionic action. However $\frac{\delta I_{M}[\psi,\bar{\psi},A]}{\delta 
A_{\mu}(x)}=-eJ^{\mu}(x)$, therefore
\begin{eqnarray}
I\left[\psi,\bar{\psi},A\right]&=&I_{F}\left[\psi,\bar{\psi}\right]+I_{G}[A]-e\int d^{n}xJ^{\mu}(x)A_{\mu}(x) 
\end{eqnarray}
\begin{eqnarray}
I_{M}\left[\psi,\bar{\psi},A\right]&=&I_{F}\left[\psi,\bar{\psi}\right]-e\int d^{n}xJ^{\mu}(x)A_{\mu}(x) 
\end{eqnarray}
Thus, evidently
\begin{equation}
-\frac{1}{e}\frac{\delta I_{M}\left[\psi,\bar{\psi},A^{\theta}\right]}{\delta A_{\mu}(x)}=-\frac{1}{e}\frac{\delta I_{M}\left[\psi,\bar{\psi},A^{\theta}\right]}{\delta A_{\mu}^{\theta}(x)}=-\frac{1}{e}\frac{\delta I_{M}\left[\psi,\bar{\psi},A\right]}{\delta A_{\mu}(x)}=J^{\mu}(x).  
\end{equation}
Since $I_{G}[A]$ is gauge invariant, which means that $\partial_{\mu}\left(-\frac{1}{e}\frac{\delta I_{G}}{\delta 
A_{\mu}(x)}\right)\equiv 0$, eq. (\ref{en-curr}) leads to
\begin{eqnarray}
\partial_{\mu}\left(-\frac{1}{e}\frac{\delta W_{eff}[A]}{\delta A_{\mu}(x)}\right)\equiv 0\Leftrightarrow\int d\theta d\psi d\bar{\psi}\partial_{\mu}J^{\mu}(x)\exp\left(iI_{en}\left[\psi,\bar{\psi},A,\theta\right]\right)\equiv 0. \label{enh-curr-cons} 
\end{eqnarray}

Eq. (\ref{enh-curr-cons}) means that the current is conserved in this version of 
HT construction, with no quantum breakdown and, thus, anomaly-free.

To finish this section, we shall analyze the \textit{classical} equations of 
motion obtained from the original abelian anomalous models
\begin{eqnarray}
\frac{\delta I\left[\psi,\bar{\psi},A_{\mu}\right]}{\delta\psi}&=&\frac{\delta I_{M}\left[\psi,\bar{\psi},A_{\mu}\right]}{\delta\psi}=0\\
\frac{\delta I\left[\psi,\bar{\psi},A_{\mu}\right]}{\delta\bar{\psi}}&=&\frac{\delta I_{M}\left[\psi,\bar{\psi},A_{\mu}\right]}{\delta\bar{\psi}}=0\\
\frac{\delta I}{\delta A_{\mu}}&=&\frac{\delta I_{M}}{\delta A_{\mu}}+\frac{\delta I_{G}}{\delta A_{\mu}}=0 
\end{eqnarray}
and compare them with those from the enhanced action $I_{en}\left[\psi,\bar{\psi}, A,\theta\right]\equiv I\left[\psi,\bar{\psi},A^{\theta}\right]$

\begin{eqnarray}
\frac{\delta I\left[\psi,\bar{\psi},A_{\mu}^{\theta}\right]}{\delta\psi}&=&\frac{\delta I_{M}\left[\psi,\bar{\psi},A_{\mu}^{\theta}\right]}{\delta\psi}=0\\
\frac{\delta I\left[\psi,\bar{\psi},A_{\mu}^{\theta}\right]}{\delta\bar{\psi}}&=&\frac{\delta I_{M}\left[\psi,\bar{\psi},A_{\mu}^{\theta}\right]}{\delta\bar{\psi}}=0\\
\frac{\delta I\left[\psi,\bar{\psi},A_{\mu}^{\theta}\right]}{\delta A_{\mu}(x)}&=&\frac{\delta I_{M}\left[\psi,\bar{\psi},A_{\mu}^{\theta}\right]}{\delta A_{\mu}(x)}+\frac{\delta I_{G}\left[A_{\mu}^{\theta}\right]}{\delta A_{\mu}(x)}=\frac{\delta I_{M}\left[\psi,\bar{\psi},A_{\mu}\right]}{\delta A_{\mu}(x)}+\frac{\delta I_{G}\left[A_{\mu}\right]}{\delta A_{\mu}(x)}=0\\
\frac{\delta I}{\delta\theta}&=&\partial_{\mu}\left(-\frac{1}{e}\frac{\delta I_{M}\left[\psi,\bar{\psi},A\right]}{\delta A_{\mu}(x)}\right)=\partial_{\mu}J^{\mu}=0 \label{theta-eq}
\end{eqnarray}

As one could see, the equation (\ref{theta-eq}) for $\theta$ is redundant, since 
it is just the current conservation law imposed by global gauge invariance. The 
equation of motion for the gauge field is the same in both theories, since it is 
gauge invariant. Finally, the equations for the fermionic fields are reducible 
one to the other by a simple redefinition of the gauge field which is nothing 
but a generic gauge transformation $A_{\mu}\rightarrow A_{\mu}'=A_{\mu}+\frac{1}{e}\partial_{\mu}\theta$ 
that does not change the other equations. Thus, classically both formulations 
are completely equivalent one to the other, and the scalar is not even noted. On 
the other hand, at quantum level, the simple original theory is anomalous, while 
the enhanced one, with the addition of the $\theta$\textit{-field} is not. 

In the next sections, we shall understand why the $\theta-field$ can be absorbed 
with no loss of physical meaning by other means.

\section{HT gauge recovering procedure applied to non-anomalous theories - The Proca Model}

As shown by the authors in the work of ref. \cite{HT2} to the case of the 
massive vector field, the procedure of turning a theory that does not exhibit 
quantum gauge symmetry into a gauge invariant one does not need 
to be restricted to the particular class of gauge anomalous models. Indeed, to 
do so, it was only necessary to consider the exponential of the effective action 
$\exp(iW[A])$, gauge transform it into $\exp(iW[A^{\theta}])$, and then to 
perform an integration over $\theta$ to obtain, finally, the exponential of the 
gauge invariant effective action $\exp(iW_{eff}[A])$. But \textit{any} action 
that does not exhibit gauge invariance could, in principle, be attached to this 
procedure. Let us reconsider, for instance, the massive vector field interacting 
with fermions, whose action is
\begin{eqnarray}
I\left[\psi,\bar{\psi},A_{\mu}\right]=I_{M}\left[\psi,\bar{\psi},A_{\mu}\right]-\frac{1}{4}\int d^{4}xF^{\mu\nu}F_{\mu\nu}+\frac{m^{2}}{2}\int d^{4}xA^{\mu}A_{\mu}. \label{Proca}
\end{eqnarray} 

Clearly, the massive term breaks gauge invariance. If we consider the quantum 
version of this model and proceed the with the HT technique, we will get
\begin{eqnarray}
\int &d{\theta}&\exp\left(iW\left[A^{\theta}\right]\right)=\int d{\theta}\exp\left(iI\left[\psi,\bar{\psi},A^{\theta}\right]\right)\nonumber\\
=\int &d{\theta}&d\psi d\bar{\psi}\exp\left(I_{M}\left[\psi,\bar{\psi},A_{\mu}^{\theta}\right]-\frac{1}{4}\int d^{4}xF^{\mu\nu}F_{\mu\nu}+\frac{m^{2}}{2}\int d^{4}xA^{\theta\mu}A_{\mu}^{\theta}   
\right)\nonumber\\
=\int &d{\theta}&d\psi^{\theta}d\bar{\psi}^{\theta}\exp\left(I_{M}\left[\psi^{\theta},\bar{\psi}^{\theta},A_{\mu}^{\theta}\right]-\frac{1}{4}\int d^{4}xF^{\mu\nu}F_{\mu\nu}+\frac{m^{2}}{2}\int d^{4}xA^{\theta\mu}A_{\mu}^{\theta}   
\right)
\end{eqnarray}
and if the theory is not anomalous, that is, if 
$d\psi d\bar{\psi}=d\psi^{\theta}d\bar{\psi}^{\theta}$, we will arrive at an enhanced model given by
\begin{equation}
\exp(iW_{eff}[A])=\int d\theta d\psi d\bar{\psi}\exp\left(iI_{en}\left[\psi,\bar{\psi},A,\theta\right]\right), \label{enh-pro} 
\end{equation} 
where
\begin{eqnarray}
I_{en}\left[\psi,\bar{\psi},A_{\mu},\theta\right]=I_{M}\left[\psi,\bar{\psi},A_{\mu}\right]+\int 
d^{4}x\left(-\frac{1}{4}F^{\mu\nu}F_{\mu\nu}\right. \nonumber\\
\left. +\frac{1}{2}\frac{m^{2}}{e^{2}}\partial^{\mu}\theta\partial_{\mu}\theta+\frac{1}{2}m^{2}A^{\mu}A_{\mu}+\frac{1}{2}\frac{m^{2}}{e}A^{\mu}\partial_{\mu}\theta\right).  
\end{eqnarray}

If we proceed integration over the gauge parameter, we will find
\begin{eqnarray}
\int &d\theta&\exp\left\{\frac{i}{2}m^{2}\int 
dx\left(\frac{2}{e}A_{\mu}\partial^{\mu}\theta+\frac{1}{e^{2}}\partial_{\mu}\theta\partial^{\mu}\theta\right)\right\}=\exp\left(-\frac{1}{2}m^{2}\int 
dxA_{\mu}\frac{\partial^{mu}\partial^{\nu}}{\Box}A_{\nu}\right)\nonumber\\
&\times&\int d\theta\exp\left\{\-i\frac{m^{2}}{2e}\int 
dx\left[\left(\frac{e}{\Box}\partial^{\mu}A_{\mu}+\theta\right)\Box\left(\frac{e}{\Box}\partial^{\nu}A_{\nu}+\theta\right)\right]\right\}.
\end{eqnarray}
Performing the change of variables $\theta\rightarrow\theta'=\theta+\frac{1}{\Box}\partial^{\mu}A_{\mu}; 
d\theta'=d\theta$, we will arrive at
\begin{equation}
 \int d\theta\exp\left\{\frac{i}{2}m^{2}\int dx\left(2A_{\mu}\partial^{\mu}\theta+\partial_{\mu}\theta\partial^{\mu}\theta\right)\right\}\sim\exp\left(-\frac{i}{2}m^{2}\int dxA_{\mu}\frac{\partial^{\mu}\partial^{\nu}}{{\Box}}A_{\nu}\right).\label{inv-Proca} 
\end{equation}
Using this result into (\ref{enh-pro}), we finally obtain
\begin{equation}
\int d\theta d\psi d\bar{\psi}\exp\left(iI_{en}\left[\psi,\bar{\psi},A_{\mu},\theta\right]\right)=\int d\psi d\bar{\psi}\exp\left(iI'\left[\psi,\bar{\psi},A_{\mu}\right]\right)  
\end{equation}
with
\begin{equation}
I'\left[\psi,\bar{\psi},A_{\mu}\right]=I_{M}\left[\psi,\bar{\psi},A_{\mu}\right]+\int d^{n}x\left\{-\frac{1}{4}F^{\mu\nu}F_{\mu\nu}+\frac{1}{2}m^{2}A_{\mu}\left(\eta^{\mu\nu}-\frac{\partial^{\mu}\partial{\nu}}{\Box}A_{\nu}\right)\right\}.  \label{inv-Proca}
\end{equation}
It is easy to see that, classically, the gauge invariant formulation of Proca model 
(\ref{inv-Proca}) may be thought as equivalent to its correlate (\ref{Proca}), 
since one is reducible to the other, with no loss of physical meaning, by the 
\textit{Lorenz} gauge choice $\partial_{\mu}A^{\mu}=0$. Therefore, this example clearly 
shows that the HT technique of inserting a quantum scalar may be used as a procedure to map a 
theory with no gauge symmetry into a gauge invariant one even in some cases 
where we are dealing with classical models.

\section{The Enhanced Formalism and the Stueckelberg Mechanism}

In the enhanced gauge invariant formalism of anomalous models, we start with a gauge invariant 
action $I_{en}\left[\psi,\bar{\psi},A,\theta\right]$, and reach an affective one 
$W_{eff}[A]$ which is also gauge invariant. However, there is an intermediate 
action $W'[A,\theta]=W\left[A^{\theta}\right]$ with no gauge symmetry. 
Nevertheless, it is obviously invariant under a kind of expanded gauge 
transformations
\begin{eqnarray}
A_{\mu}&\rightarrow& A_{\mu}+\frac{1}{e}\partial_{\mu}\Lambda(x)\nonumber\\
\theta&\rightarrow&\theta-\Lambda(x) \label{Pauli'}
\end{eqnarray}

It means that we can set $\theta(x)=constant$ by a simple gauge choice and get back to the original formalism. In other words, classically, $\theta$ is not noted, but must exist and be quantized in order to provide an anomaly-free model. In section 4 we saw that the classical equations of motion of the enhanced version of anomalous models are reducible to those of the original one by a simple redefinition of the gauge boson. By the modified gauge symmetry (\ref{Pauli'}) above, thus, it simple means a gauge choice where the scalar is set constant.
On the other hand, the pure enhanced Proca model, which is also invariant under 
(\ref{Pauli'}), is described by
\begin{equation}
I_{P}[A,\theta]=\int d^{n}x\left(-\frac{1}{4}F^{\mu\nu}F_{\mu\nu}+\frac{1}{2}\frac{m^{2}}{e^{2}}\partial^{\mu}\theta\partial_{\mu}\theta+\frac{1}{2}m^{2}A^{\mu}A_{\mu}+\frac{m^2}{e}A^{\mu}\partial_{\mu}\theta\right).  
\end{equation}
If we simply redefine the $\theta-field$ by a multiplicative constant
\begin{equation}
B(x)\equiv\frac{m}{e}\theta(x),  
\end{equation}
then we will just find the Stueckelberg action \cite{Stueck}
\begin{equation}
I_{Stueck}[A,B]=\int d^{n}x\left\{-\frac{1}{4}F^{\mu\nu}F_{\mu\nu}+\frac{1}{2}(mA^{\mu}+\partial^{\mu}B)(mA_{\mu}+\partial_{\mu}B)\right\},  
\end{equation}
and (\ref{Pauli'}) becomes Pauli's gauge transformations \cite{Pauli}
\begin{eqnarray}
A_{\mu}&\rightarrow& A_{\mu}+\partial_{\mu}\Lambda(x)\nonumber\\
B&\rightarrow&B-m\Lambda(x). \label{Pauli}
\end{eqnarray}

Therefore, we see that the HT procedure, using the enhanced form in 
the case of abelian models, is in closed connection with the Stueckelberg 
mechanism \textit{before} the integration over the scalar, and may be viewed as its generalization, whose prescription is to 
attach a gradient of a scalar added up to every gauge boson.

The advantage of the Stueckelberg massive abelian model, which coincides exactly 
with the HT gauge invariant procedure applied to the Proca model before the 
integration over the extra degree of freedom, is that it was rigorously proved 
to be renormalizable and unitary \cite{Schroer}.

We started by the integration over what we called the gauge parameter, but now we can 
reinterpret it by saying that it is not the gauge parameter which is actually 
integrated, but the Stueckelberg scalar. The procedure presented above 
shows that  the Stueckelberg field can appear quite naturally, by going to 
the quantum level, and performing analytical manipulations over the functional
integral that actually reveal it, instead of using the rather artificial Stueckelberg
trick by adding the extra degree of freedom classically by hand. At the end, the 
Stueckelberg trick may be justified under this approach. We can interpret the 
Stueckelberg scalar as a hidden field, which is physically non observable at 
this point, but becomes necessary whenever we deal with gauge symmetry breaking and, as the mentioned examples,  
want to be able to provide a gauge anomaly-free theory as well as a renormalizable massive vector model.

\section{Conclusion}

Revisiting a procedure to transform effective actions of anomalous generic 
models into gauge invariant ones, built in the last century, it was found that 
it can be more fruitful than it might have seem to be at a first sight. Indeed, the HT procedure is not only able to map an anomalous model into a gauge invariant one, but it may also be able to remove abelian gauge anomalies, which simply disappear when the $\theta-field$ 
is introduced into the theory by gauge transforming the gauge field. Moreover, 
it provides a bridge between the gauge invariant formulation of gauge 
anomalous models and a generalization of the Stueckelberg procedure, where the 
$\theta-field$, identified as the Stueckelberg scalar, may be present together 
with the gauge field in any abelian theory, instead of being present only in the 
particular case of the massive vector model. The Stueckelberg mechanism was also 
extended to non-abelian gauge models, as can be seen in \cite{Banerjee}. Perhaps the HT procedure, which was originally proposed for Yang-Mills models \cite{HT}, may also be linked to this extension in the same way. 

On the other hand, such discussion may raise a paradox: If one formalism is 
mapped into another one by simple manipulations over the functional integral, 
which would suggest that both formalisms are equivalent, how, in the anomalous 
case, one might present current conservation breakdown while the other has it 
conserved? As we have seen, the original formalism is anomalous, which would 
mean that it is closer to the standard formalism, that preserves 
its anomaly, then to the enhanced one. In this sense, one might ask which of the 
two gauge invariant models is equivalent to the original one. This question 
may be partially answered in \cite{GRT} where it was shown that the original anomalous formalism has the expectation value of the anomaly cancelled out when one goes to the full quantum theory, \textit{i. e.}, the one with the gauge fields also quantized. 
Work is in progress to clarify this question in more detail.

The relevance of the Stueckelberg mechanism is that it is able to deal with gauge symmetry breaking and, since it is renormalizable, it provides a mechanism alternative to the Higgs 
\cite{Higgs}. Moreover, it can be recovered in a rather singular limit of the 
Higgs mechanism \cite{Allen}. Therefore, perhaps the uncovered hidden 
scalar field might be regarded as an inheritance of the Higgs mechanism at lower 
energies. In revealing a generalization of the Stueckelberg mechanism, we saw 
that it is also able to provide a gauge anomaly-free model. On the other hand, 
it is well known, for the simplest case of the anomalous Jackiw-Rajaraman model, 
that there is an alternative mass-generation mechanism to the gauge boson from 
quantum corrections of anomalous \textit{2-D} chiral fermions \cite{JR}. Perhaps 
it is not mere coincidence that a breaking in the gauge symmetry in both cases 
is related to vector boson mass generation, and it may be recovered by an 
introduction of a scalar.

We can point out that, besides the correspondence between the HT procedure
and the Stueckelberg mechanism, this technique might be generalized to other
kind of symmetries, although it has been remarked that it may not able to deal with chiral anomalies, for example \cite{GRT}. It is well known that the Stueckelberg trick can also be used to restore
gravitational gauge symmetry to deal with massive gravity models, for instance \cite{KH}. One might ask whether
it would be justified by a kind of the prescription presented above, as it was shown for vector models. Finally, 
yet with the gravitational example, this procedure might be a road to cancel the gravitational anomaly, presented by
the famous work of Witten and Gaum\'{e} \cite{WG}. In this sense, it was shown that the Hawking 
radiation can be explained by the addition of chiral fermions at the boundary of 
a black hole, that cancels the gravitational anomaly \cite{RW}. Perhaps it could happen in 
a more natural way, using the HT prescription adapted to the gravitational 
case, by substituting the anomalous chiral fermions by Stueckelberg fields.

\begin{acknowledgements}
I would like to thank Prof. A. J. Accioly and Prof. Jos\'{e} A. Helay\"{e}l Neto for further revisions. I also thank my colleagues Ricardo Kullock and Ricardo Scherer for constructive discussions about the subject of this work. Finally, I am very grateful to Professor Abhay Ashtekar and the \textit{Institute for Gravitation and the Cosmos} (Penn State University) for the hospitality. This work was financially supported by CNPq through the Csf program (Ci\^{e}ncia sem Fronteiras) from Brazil and in part by the NSF grant PHY 1205388 and the Eberly research funds of Penn State University.
\end{acknowledgements}

\end{document}